\documentclass{iopconfser}
\usepackage[dvipdfmx]{graphicx}
\usepackage{bm,latexsym,amsmath,amssymb,amsfonts} 
\DeclareMathAlphabet{\mathpzc}{OT1}{pzc}{m}{it}
\usepackage[normalem]{ulem}
\usepackage[breaklinks=true]{hyperref}
\usepackage{xcolor}
\usepackage{wasysym}
\usepackage[mathscr]{eucal}
\usepackage{multirow}
\usepackage{enumitem}
\usepackage{tabularray}
\usepackage{tablefootnote}
\usepackage{longtable}
\usepackage{float}
\usepackage{booktabs}

\begin{document}

\title{Shadow images of regular black hole with finite boundary}

\author{M.~F.~Fauzi$^{1\ast}$, H.~S.~Ramadhan$^{1}$ and A.~Sulaksono$^{1}$}

\affil{$^1$Departemen Fisika, FMIPA, Universitas Indonesia, Depok 16424, Indonesia}

\email{$^\ast$muhammad.fahmi31@ui.ac.id}

\begin{abstract}
Regular black hole is one of the bottom-up solutions designed to eliminate the singularity at the center of black holes. Its horizonless solution has gained interest recently to model ultracompact star. Despite interesting, this proposal is problematic due to the absence of a well-defined boundary. In this work, we introduce a novel regular black hole model inspired by the Hayward black hole, incorporating additional terms to define a clear and well-defined `surface' radius $R$. We analyze the null geodesics around the object, both horizonful and horizonless configurations, by studying the photon effective potential. We further simulate the shadow images of the object surrounded by a thin accretion disk. Our results indicate that for $R > 3M$ the horizonfull shadow differs slightly from that of a Schwarzschild black hole. In the horizonless configuration, we identify distinct inner light ring structures near the central region of the shadow image, which differ from those observed in horizonless Hayward black holes.
\end{abstract}

\section{Introduction}
One of the most intriguing problem in classical general relativity is the prediction of the existence of black holes, {\it i.e.,} singularity hidden beneath the horizons~\cite{Hobson:2006se}. All canonical black holes are plagued by singularity. Enormous proposed solutions to resolve the singularity has been presented in the literature (for a short discussion, see Ref.~\cite{Cadoni:2022chn}), but so far the matter is still far from settled.

A classical bottom-up approach to resolving singularities in black holes is through the concept of regular black holes (RBHs). An RBH metric eliminates singularities, ensuring that all invariants remain finite everywhere. The simplest method to construct an RBH involves replacing the constant mass parameter $M$ in the metric with a position-dependent mass function $m(r)$ \cite{Lan:2023cvz}:
\begin{equation}
    \lim_{r\to0} \frac{m(r)}{r} = 0
\end{equation}
which asymptotically satisfies $\lim_{r\to\infty} m(r) = M$. One approach to determine the source of the mass function is through de Sitter vacuum, where the energy momentum tensor is in the perfect fluid form with the de Sitter equation of state (EoS): $p=-\epsilon$, where $p$ and $\epsilon$ are the radial pressure and energy density, respectively. Since the energy density is considered to be always positive, the negative pressure implies that the spacetime has a repulsive core, which prevents the mass distribution from further collapse into a single point, thus avoiding the formation of singularity. It is always possible to obtain a horizonless RBH spacetime known as the ``star" solution~\cite{Carballo-Rubio:2022nuj}, but it has a problematic implementation since most of them do not have a well-defined radius. With the de Sitter EoS, this implies imaginary speed of sound everywhere.

Remarkable developement of the Event Horizon Telescope (EHT) \cite{EventHorizonTelescope:2024xos} brings extensive studies to the shadow images of black hole models and other ultracompact object, e.g., in Refs.~\cite{Carballo-Rubio:2022aed,Meng:2023htc,Rosa:2024bqv,Rosa:2024eva}. With higher resolution, EHT may provides us more insights on the nature of black holes \cite{Eichhorn:2022fcl}, and possibly find any deviation from current theories that might be explainable through regular black holes.

In this paper, we propose a finite boundary model of RBH and generate its shadow images. This paper is organized as follows. In Sec.~\ref{sec. rbh with finite}, we start the construction by modifiying the energy density of Hayward's RBH, introducing an additional term that defines a `surface' radius $R$. Form the model, in Sec.~\ref{sec. shadow images} we generate the images of the black hole surrounded by thin accretion disk, as well as the photon trajectories around the spacetime.

\section{Regular black holes with finite boundary}
\label{sec. rbh with finite}

We adopt the spacetime metric in form of
\begin{equation}
    ds^2 = -A(r) dt^2 + \frac{1}{A(r)} dr^2 + r^2 d\Omega^2,
\end{equation}
where $d\Omega^2 = \sin^2\theta d\theta ^2 + d\phi^2$.

Here is a revised version of the sentence:

To eliminate the singularity at the center of a black hole, the RBH approach typically introduces a new physical parameter that avoids the divergence at $r = 0$. This parameter could be a charge produced by nonlinear electrodynamics or a new physical scale, particularly related to the quantum nature at extreme curvature \cite{Lan:2023cvz, Cadoni:2022chn}. One of the simplest models of RBHs is the Hayward model \cite{Hayward:2005gi,Cadoni:2022chn}, described by
\begin{equation}
    A(r) = 1 - \frac{2m(r)}{r}, \qquad m(r) \equiv \frac{Mr^3}{r^3 + \ell^3},
    \label{eq. ansatz SSS}
\end{equation}
where $\ell$ is a new parameter that gives rise to a de Sitter-like spacetime near $r = 0$. If one considers that the spacetime is sourced by the energy-momentum tensor of a perfect fluid with a de Sitter equation of state $p = -\epsilon$, the energy density can be written as
\begin{equation}
    \epsilon(r) = \frac{3}{4\pi} \frac{M\ell^3}{\left(r^3 + \ell ^3\right)^2}
\end{equation}
The Hayward spacetime structure is primarily determined by the ratio \(\alpha = 2M/\ell\). There are three types of configurations: horizonful (\(\alpha > \alpha_c\)), extremal (\(\alpha = \alpha_c\)), and horizonless (\(\alpha < \alpha_c\)). The value of \(\alpha_c\) is determined by the condition \(A(r) = \partial_r A(r) = 0\), and it is known to be \(\alpha_c = 3/\sqrt[3]{4}\), while the extremal horizon location \(r_H\) is at \(r_H = 4M/3\).

In this study, we introduce a new term in the energy density to define a finite boundary $R$,
\begin{equation}
    \epsilon(r) = 
    \begin{cases}
        \frac{3}{4\pi} \frac{\kappa M\ell^3}{\left(r^3 + \ell ^3\right)^2} \left(1 - \left(\frac{r}{R}\right)^n\right), & 0>r>R\\
        0, & r>R
    \end{cases}
\end{equation}
and it gives rise to the mass profile
\begin{equation}
    m(r) = 
    \begin{cases}
         \kappa Mr^3\left[\frac{1}{r^3 + \ell^3} - \frac{3\left(\frac{r}{R}\right)^n {}_2F_1\left(2,\frac{n}{3}+1,\frac{n}{3}+2,-\frac{r^3}{\ell^3}\right)}{l^3(3+n)}\right], & 0\leq r \leq R\\
         M, & r>R
     \end{cases}
     \label{eq. mass profile R}
\end{equation}
where $_2F_1$ is a hypergeometric function, and
\begin{equation}
    \kappa = \left[R^3 \left(\frac{1}{\ell^3+R^3}-\frac{3 \, _2F_1\left(2,\frac{n}{3}+1;\frac{n}{3}+2;-\frac{R^3}{\ell^3}\right)}{\ell^3 (n+3)}\right)\right]^{-1}.
\end{equation}
For the sake of simplicity, we will use $n=3$, so that the hypergeometric function can be represented by logarithmic functions.

\begin{figure}[htbp!]
    \centering
    \includegraphics[width=0.6\linewidth]{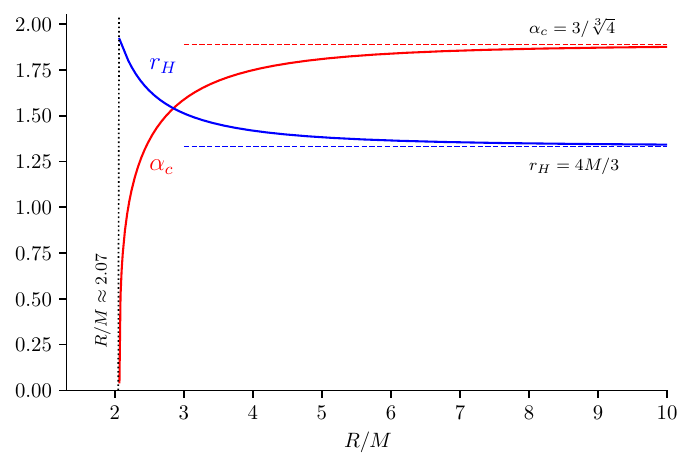}
    \caption{Critical value of $\alpha$ for different values of $R/M$.}
    \label{fig. alpha critical}
\end{figure}

As a consequence of this additional term, \(\alpha_c\) now depends on the value of \(R/M\), as shown in Fig.~\ref{fig. alpha critical}. In the large \(R/M\) limit, both \(\alpha_c\) and \(r_H\) approach the original Hayward spacetime values of \(3/\sqrt[3]{4}\) and \(4M/3\), respectively. \(\alpha_c\) does not exist for all values of \(R/M\); instead, there is a lower limit at \(R/M \approx 2.07\), below which \(\alpha_c\) does not have a real value. Below this threshold, the spacetime always contains two horizons. This is expected, as the Schwarzschild black hole horizon is located at \(r/M = 2\).

\section{Photon effective potential and shadow images}\label{sec. shadow images}
Photon geodesics on a spacetime manifold can be calculated by solving the geodesic equation \cite{Hobson:2006se},
\begin{equation}
    \frac{du^\mu}{d\tau} + \Gamma^{\mu}_{\alpha\beta} \frac{dx^\alpha}{d\tau} \frac{dx^\beta}{d\tau} = 0
\end{equation}
where \(x^\mu = [t, r, \theta, \phi]\) is the photon four-vector, \(u^\mu = \frac{dx^\mu}{d\tau}\) is the photon four-velocity, and \(\tau\) is an affine parameter. In a static and spherically symmetric spacetime of the form given by Eq.~\eqref{eq. ansatz SSS}, we can analyze the radial behavior of the photon trajectory by studying its effective potential (see, e.g., Refs.~\cite{Wang:2023vcv, Olmo:2023lil, Cao:2023par}),
\begin{equation}
    V(r) = \frac{A(r)}{r^2}.
\end{equation}
One interesting feature of ultracompact spacetimes is the existence of circular photon orbits, known as the photon sphere. The photon sphere radius \(r_{ps}\) can be found by demanding \(\frac{dV}{dr} = 0\), and in the case of the Schwarzschild metric, the photon sphere is located at \(r_{ps} = 3M\) \cite{Hobson:2006se}. As can be seen from Eq.~\eqref{eq. mass profile R}, our proposed spacetime metric reduces to the Schwarzschild metric for \(r > R\). Hence, if \(R \leq 3M\), our regular black hole model would also have a photon sphere at \(r_{ps} = 3M\).

There are two interesting types of cases: horizonful configuration with \(R > 3M\) and horizonless configuration with \(R \leq 3M\). In the first case, the photon sphere radius will be slightly modified since the spacetime is non-Schwarzschild at \(R = 3M\), resulting in a different shadow radius compared to the Schwarzschild black hole. In the second case, a new stable photon orbit will always form and be accessible\textemdash even for arbitrarily small values of \(\alpha\)\textemdash since no horizon exists and the spacetime at the center is flat. Hence, assuming that the object is surrounded by a thin accretion disk, multiple ring-shaped secondary images will be produced in the central part of the optical image. However, in this study, we will focus on comparing how the radius of the boundary influences the shadow images, so we use the same \(R/M\) range for both configurations.

\begin{figure}[htbp!]
    \centering
    \includegraphics[width=0.8\linewidth]{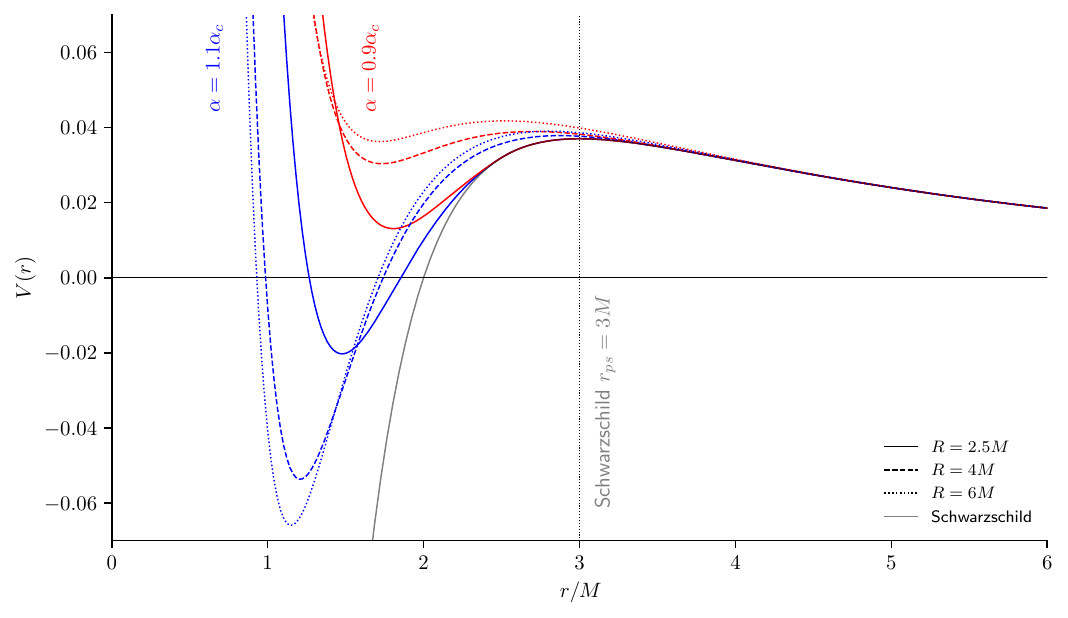}
    \caption{Photon effective potential for several values of $R/M$ with $\alpha = 1.1\alpha_c$ (blue) and $\alpha=0.9\alpha_c$ (red).}
    \label{fig. effective potential}
\end{figure}

We compare the effective potential for both horizonful and horizonless configurations with \(\alpha = 1.1\alpha_c\) and \(\alpha = 0.9\alpha_c\), respectively. For both cases, we vary \(R/M = [2.5, 4, 6]\). Without loss of generality, we choose \(\ell = 1\). The results are shown in Fig.~\ref{fig. effective potential}. It can be seen that the horizonful and horizonless configurations exhibit opposite behaviors: smaller \(R/M\) values deepen the `potential well' in the horizonless configuration, while the opposite occurs in the horizonful configuration. It is also confirmed that for \(R/M \leq 3\), the local maximum of \(V(r)\) is located at \(R/M = 3\), and \(R/M > 3\) shifts it inward, producing smaller photon sphere radius. We observe a local minimum of \(V(r) > 0\) in the horizonless configuration, indicating that a stable photon sphere exists inside the boundary.

\subsection{Photon trajectories} \label{sec. photon trajectories}
Null geodesics or photon trajectories can be computed by integrating \cite{Rosa:2023hfm,Wang:2023vcv,Olmo:2023lil}
\begin{equation}
    \frac{d\phi}{dr} = \frac{1}{r^2}\sqrt{\frac{1}{b^{-2} - V(r)}}
\end{equation}
where \(b\) is the impact parameter of a photon. The integration is performed numerically using backward integration from an observer located at numerical infinity. The process is carried out in two parts: inward and outward. First, we integrate the null geodesic inward by setting a negative step size. Once the integration reaches the turning point where \(V(r) \geq 1/b^2\), the direction of integration is reversed, and the step size is set to a positive value.

\begin{figure}[htbp!]
    \centering
    \includegraphics[width=1\linewidth]{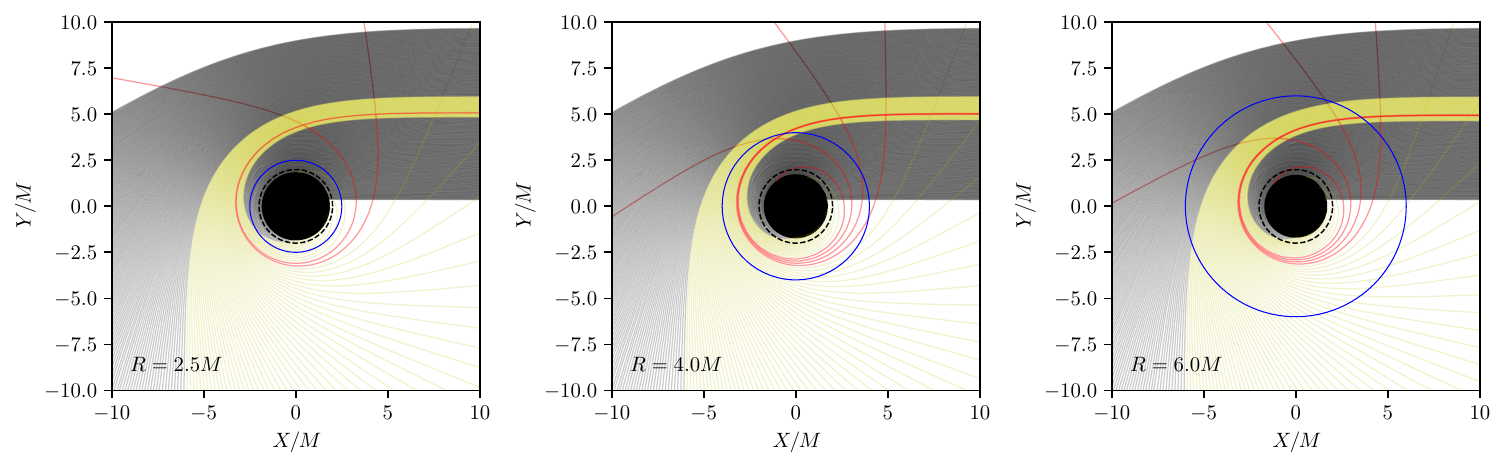}
    \caption{Photon trajectories around the horizonful configuration. The direct emission, lensed emission, and photon ring trajectories are represented by black, yellow, and red lines, respectively. The blue circle indicates the boundary radius \(R/M\), and the dashed black circle represents the Schwarzschild radius. The solid black circle at the center denotes the horizon.}
    \label{fig. trajectory horizonful}
\end{figure}

\begin{figure}[htbp!]
    \centering
    \includegraphics[width=1\linewidth]{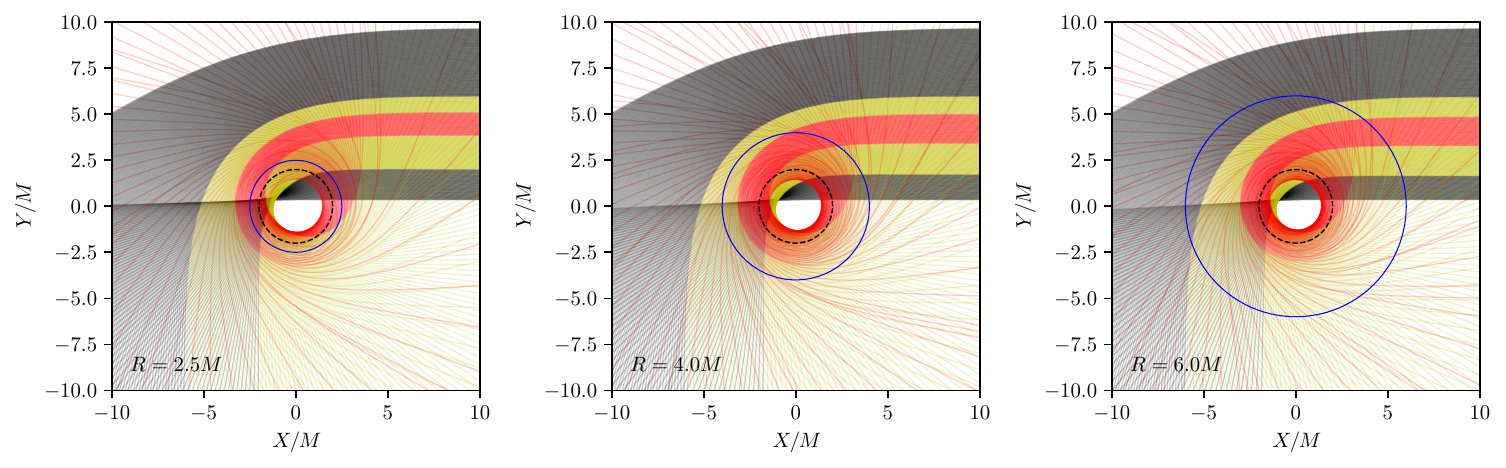}
    \caption{Same description as Fig.~\ref{fig. trajectory horizonless} for the horizonless configuration. The black solid circle is absent because there is no horizon in the spacetime.}
    \label{fig. trajectory horizonless}
\end{figure}

The photon trajectory is classified into three types based on the number of orbits, \(n_r\) (see, e.g., Ref.~\cite{Meng:2023htc}): \textit{direct emission} (\(n_r \leq 3/4\)), \textit{lensed emission} (\(3/4 < n_r < 5/4\)), and \textit{photon ring} (\(n_r \geq 5/4\)). We show the photon trajectories for the horizonless and horizonful configurations in Figs.~\ref{fig. trajectory horizonful} and \ref{fig. trajectory horizonless}, respectively.

The event horizon acts as a stopping condition for light trajectories. Based on Fig.~\ref{fig. trajectory horizonful}, our model produces a smaller horizon radius than the Schwarzschild radius, and this radius decreases with increasing \(R/M\). However, since the photon sphere radius remains the same as the Schwarzschild radius for \(R/M \leq 3\), photons with an impact parameter greater than the critical Schwarzschild value \(b_c = 3\sqrt{3}M\) will eventually fall into the horizon, resulting in the same shadow radius as that of the Schwarzschild black hole. For \(R/M > 3\), the critical impact parameter changes and is slightly "widened," producing the photon ring trajectories shown by the red lines in Fig.~\ref{fig. trajectory horizonful}.

Photon trajectories become slightly more interesting in the horizonless configuration. The absence of a horizon allows photons to potentially pass through the center, resulting in a wider range of impact parameters for lensed emission and photon ring trajectories. As a consequence, several additional secondary images would appear in the inner part of the optical image, and the system loses its "shadow" feature.

\subsection{Accretion disk and ray tracing}
To generate the optical images, we use ray tracing methods by calculating null geodesics from the observer to a light source near the object, which, in this case, is a thin accretion disk. The intensity detected along the null geodesic is then recorded onto a screen pixel to generate the images.

The thin accretion disk used in this study spans the equatorial plane, and its intensity profile is described by the well-known Gralla-Lupsasca-Marrone (GLM) model \cite{Gralla:2020srx},
\begin{equation}
    I(r) = \frac{\exp\left\{-\frac{1}{2}\left[\gamma + \operatorname{arcsinh}{\left(\frac{r-\mu}{\sigma}\right)}\right]^2\right\}}{\sqrt{\left(r-\mu\right)^2+\sigma^2}},
	\label{eq. intensity profile glm}
\end{equation}
where $\gamma$, $\mu$, and $\sigma$ are parameters that determine the shape of the intensity profile. We choose $\gamma = -2$, $\mu = R_{ISCO} \equiv 6M$, and $\sigma = M/4$, which is a commonly used choice in the literature \cite{Rosa:2024bqv,Rosa:2023hfm}.

\begin{figure}[htbp!]
    \centering
    \includegraphics[width=0.9\linewidth]{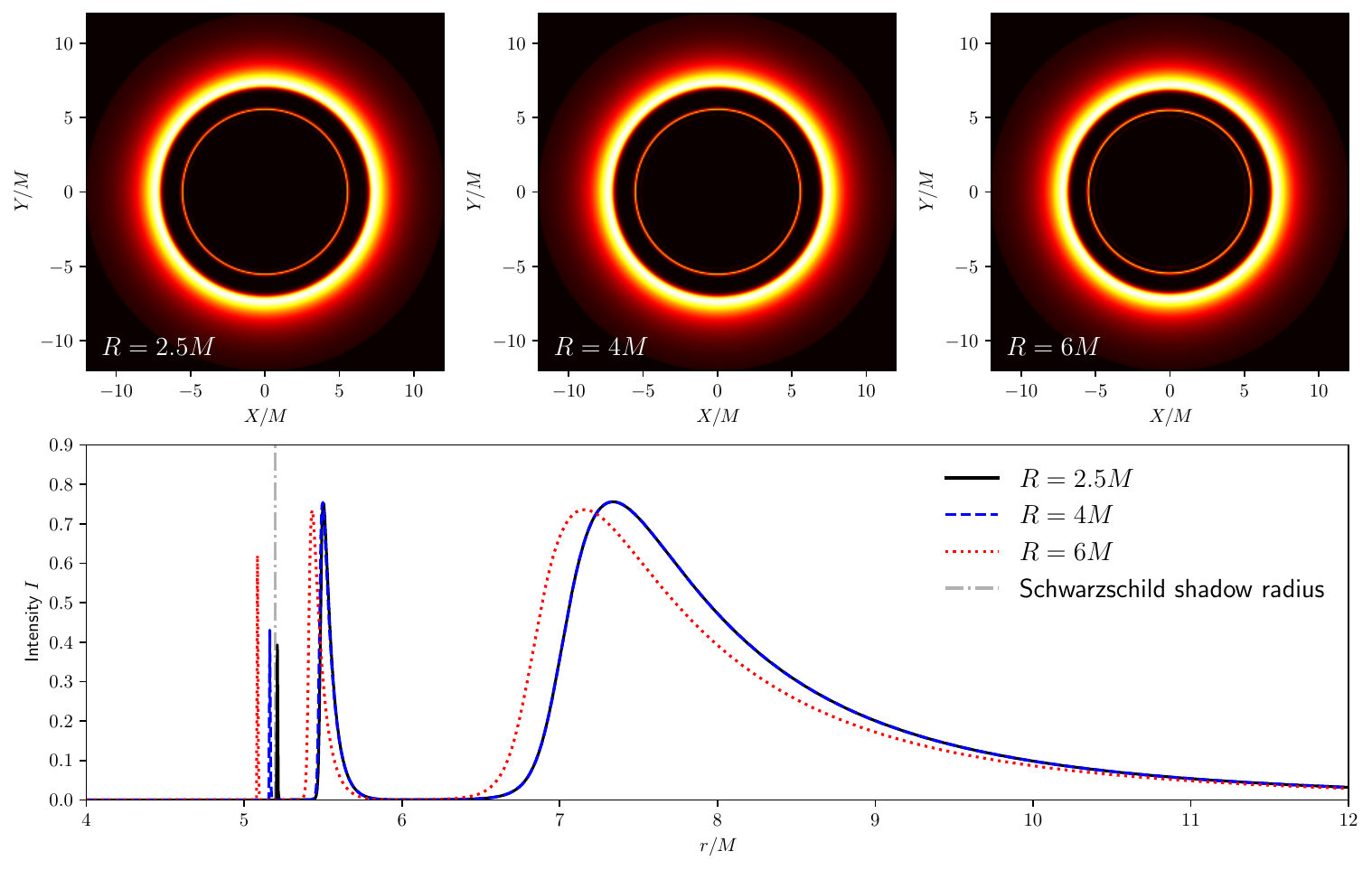}
    \caption{(Top row) Shadow images of the horizonful configuration for several values of \(R/M\). (Bottom row) Intensity cross-sections of each shadow image. The gray dotted line indicates the shadow radius of the Schwarzschild black hole, and the thin colored lines (black, blue, and red) represent the outer shadow radius for each configuration.}
    \label{fig. images horizonful}
\end{figure}

\begin{figure}[htbp!]
    \centering
    \includegraphics[width=0.9\linewidth]{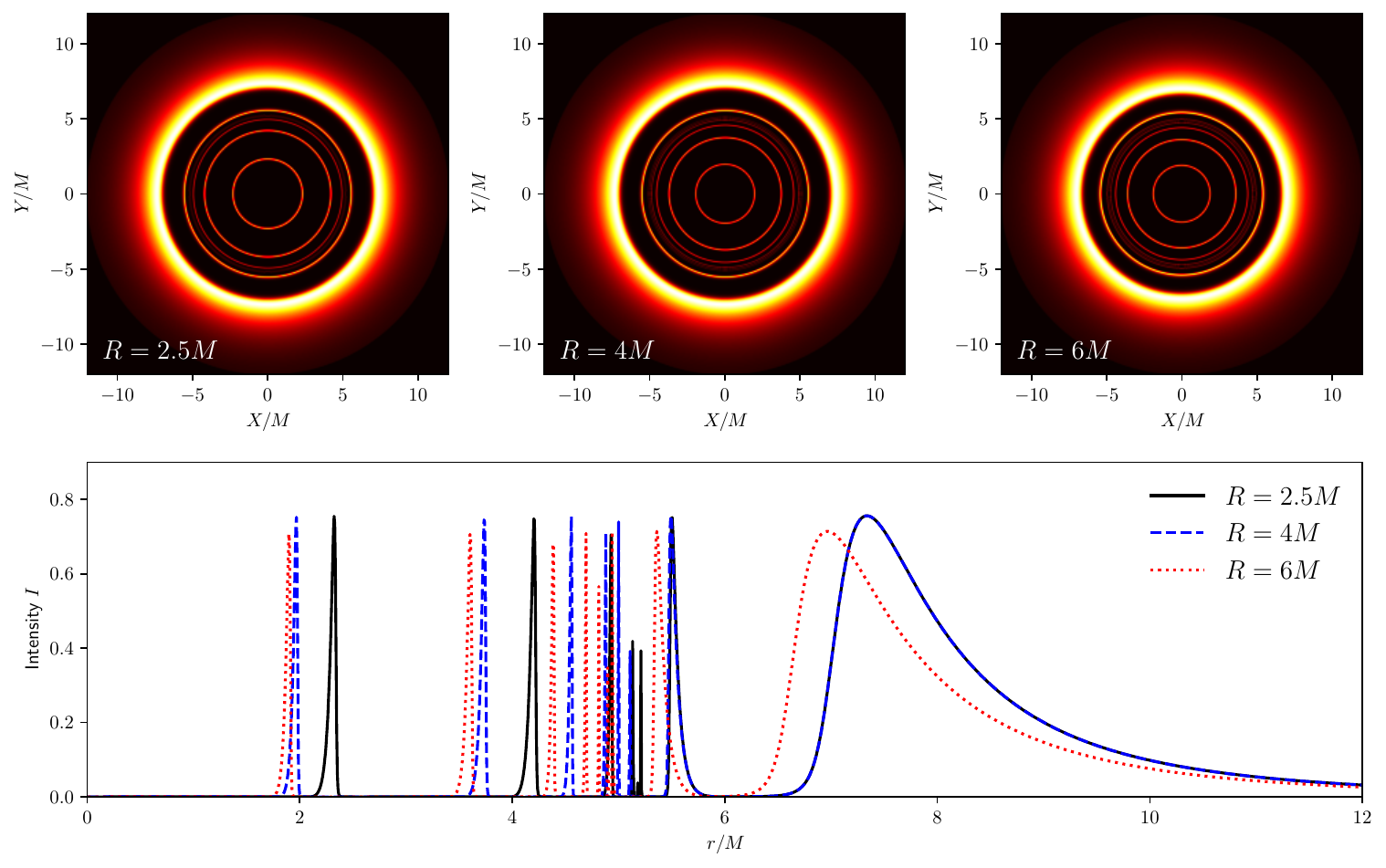}
    \caption{(Top row) Optical images of the horizonless configuration for several values of \(R/M\). (Bottom row) Intensity cross-sections of each shadow image.}
    \label{fig. images horizonless}
\end{figure}

We generate and examine the images using the parameters mentioned earlier in Sec.~\ref{sec. photon trajectories}. The results are shown in Figs.~\ref{fig. images horizonful} and \ref{fig. images horizonless} for the horizonful and horizonless configurations, respectively.

There are no noticeable differences in the shadow images of the horizonful configurations, especially for low values of \(R/M\). Only a slight modification to the shadow radius (shown as thin peak lines on the observed intensity plot, which gets smaller with increasing \(R/M\)) appears at \(R/M = 4\). At \(R/M = 6\), the direct image of the accretion disk is also shifted inward. Looking at the trajectories in Fig.~\ref{fig. trajectory horizonful}, the shifted images in the \(R/M = 6\) configuration are caused by the wider modified spacetime, which affects the entire range of lensed trajectories, as well as a few direct emission trajectories.

The horizonless spacetime produces entirely different images compared to the horizonful configuration, as predicted previously. Several ring-shaped secondary images appear in the central region, generally forming three wide rings. Thin rings are produced between the first and second outer wide rings, caused by the "chaotic" photon ring trajectories shown by the red lines in Fig.~\ref{fig. trajectory horizonless}. Increasing the value of \(R/M\) causes the inner ring to shift inward, and it exhibits similar behavior to the horizonful configuration in the direct emission images.

\section{Conclusion}
In this paper, we primarily study the image features of regular black hole spacetimes with finite boundaries, specifically focusing on modifications to Hayward’s regular black hole spacetime. The introduction of a finite boundary alters the structural properties of the spacetime, with the critical extremum configuration now depending on the boundary radius \(R\). In the limit \(R/M \to \infty\), the critical value approaches the original Hayward value of \(3/\sqrt[3]{4}\). Noticeable differences in the shadow images appear in spacetimes with horizons at large boundary radii, where both the direct and lensed images shift inward, and the shadow radius becomes smaller. In horizonless spacetimes, we observe a similar ring-shaped secondary image pattern\textemdash consisting of three wide rings and several thin rings between the first and second outer rings\textemdash across all boundary radius configurations, with slight variations in the sizes of the rings.

\section*{Acknowledgements}
We would like to thank Rafiqy Ramadhan for his valuable input during the early stages of our discussions. A. Sulaksono is funded by PUTI grant 2024-2025 No. NKB-380/UN2.RST/HKP.05.00/2024.

\end{document}